\documentclass[twocolumn, showpacs, longbibliography, prx, aps, amsmath, amssymb,superscriptaddress]{revtex4-2}
\usepackage{amsmath}
\usepackage{amsfonts}
\usepackage{graphicx}
\usepackage[colorlinks,linkcolor=blue,citecolor=blue,urlcolor=blue]{hyperref}
\usepackage{float}
\usepackage[normalem]{ulem}
\usepackage[left,columnwise, running]{lineno}
\usepackage[dvipsnames]{xcolor}
\begin{document}

\title{Continuous-Wave Nonlinear Polarization Control \\and Signatures of Criticality in a Perovskite Cavity}

\author{G. Keijsers}
\affiliation {Center for Nanophotonics, AMOLF, Science Park 104, 1098 XG Amsterdam, The Netherlands}

\author{R. M. de Boer}
\affiliation {Center for Nanophotonics, AMOLF, Science Park 104, 1098 XG Amsterdam, The Netherlands}

\author{B. Verdonschot}
\affiliation {Center for Nanophotonics, AMOLF, Science Park 104, 1098 XG Amsterdam, The Netherlands}

\author{K. J. H. Peters}
\affiliation {Center for Nanophotonics, AMOLF, Science Park 104, 1098 XG Amsterdam, The Netherlands}

\author{Z. Geng}
\affiliation {Center for Nanophotonics, AMOLF, Science Park 104, 1098 XG Amsterdam, The Netherlands}

\author{S. R. K. Rodriguez}  \email{s.rodriguez@amolf.nl}
\affiliation {Center for Nanophotonics, AMOLF, Science Park 104, 1098 XG Amsterdam, The Netherlands}

\begin{abstract}
Halide perovskites have emerged as promising photonic materials for fundamental physics studies and technological applications. Their potential for nonlinear optics has also drawn great interest recently; yet, to date, continuous-wave (CW) nonlinearities have remained elusive. Here we demonstrate CW nonlinear phenomena in a CsPbBr$_3$ perovskite cavity. We first demonstrate optical bistability --- the hallmark of single-mode coherent nonlinear optics. Next we exploit the interplay of nonlinearity and  birefringence to demonstrate nonlinear control over the polarization of light.  Finally, by measuring the optical hysteresis of our cavity as a function of temperature, we find  a dramatic enhancement of the nonlinearity around 65 K.  This enhancement is indicative of a phase transition in CsPbBr$_3$. Our results position CsPbBr$_3$ cavities as an exceptional platform for nonlinear optics, offering strong CW nonlinearity and birefringence which are furthermore tunable. In addition, our approach to uncover a phase transition of matter via optical hysteresis measurements is promising for exploring strongly correlated states of light-matter systems.
\end{abstract}

\date{\today}
\maketitle

Halide perovskites (HP) have inspired major research developments in optoelectronics~\cite{Stranks15} and photonics~\cite{Sutherland16, Makarov19} over the past decade. Among their many interesting features, HPs host large-binding-energy excitons suitable for strong coupling to optical modes at room temperature~\cite{Su21c}. This strong coupling results in part-light part-matter quasi-particles known as polaritons, which can interact via their exciton part and enhance optical nonlinearities~\cite{Fieramosca19}. Experiments under non-resonant or pulsed excitation have revealed nonlinear phenomena such as condensation~\cite{Su18, Bao19, Su20, Tao22}, optical switching~\cite{Feng21, Su21}, parametric scattering~\cite{Wu21, Wu21b}, and superfluidity~\cite{Peng22} of polaritons in HPs. However, under coherent CW driving, HPs typically bleach at power densities below nonlinear thresholds.

Among the various nonlinear effects pursued with HPs~\cite{Zhou20} and other emergent materials~\cite{Sanvitto16}, those involving an intensity-dependent refractive index rank highly. The Kerr effect, for example, involves an instantaneous refractive index change and is key to fascinating phenomena such as superfluidity~\cite{Carusotto13}, photon blockade~\cite{Chang14}, and nonlinear polarization control~\cite{Paraiso10, Moroney22}. The related thermo-optical nonlinearity, which involves a non-instantaneous refractive index change, has recently drawn interest for optical switching and isolation~\cite{King24, Cotrufo24}. In polariton systems, polariton-polariton interactions mediate Kerr nonlinearities~\cite{Carusotto13}, while polariton-exciton interactions yield a non-instantaneous refractive index change~\cite{Estrecho19, Schmidt19}. Regardless of the nonlinear response time, CW operation is crucial for many fundamental studies and applications. Otherwise, a steady state cannot be reached and functionalities are limited. Imagine, for example, an optical memory with a lifetime that is as short as the sub-picosecond pulse pumping it, or an isolator that burns if the incident light is CW or pulsed with a repetition rate exceeding the cooling rate (typically MHz) of its constituent materials.

In this work we demonstrate CW nonlinear phenomena in a CsPbBr$_3$ perovskite polariton cavity. Our experiments, at 5 K,  evidence optical bistability when probing a single mode, and signatures of tristability when probing two orthogonally-polarized modes nonlinearly coupled. Leveraging the nonlinearity of the two-mode system, we demonstrate full polarization rotation by controlling the laser-cavity detuning. We furthermore explore the temperature-dependent optical hysteresis of our cavity. Our experiments reveal a dramatic enhancement of the nonlinearity strength around 65 K, indicative of a phase transition in CsPbBr$_3$. \\

\begin{figure*}[!]
	\includegraphics[width=\textwidth]{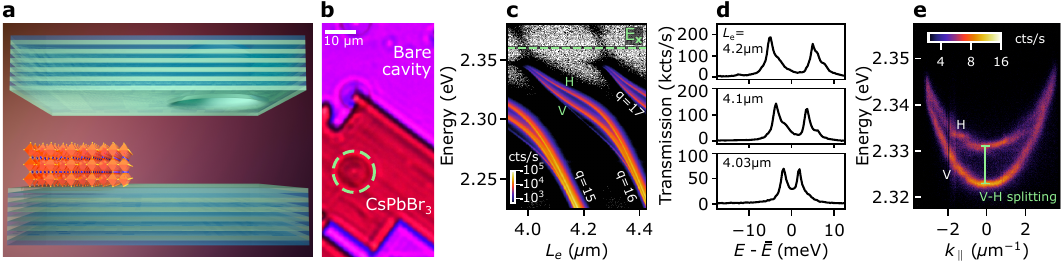}
	\caption{\label{fig:1}\textbf{Tunable birefringent perovskite cavity.}
    (a) Sketch of the system under study: A cavity made by two DBR mirrors and hosting a CsPbBr$_3$ perovskite crystal. We switch between a planar and a plano-concave cavity by translating the top mirror, which contains a micron-scale concave mirror.
    (b) White light transmission image showing the CsPbBr$_3$ crystal in red, empty regions of the cavity in pink, and a concave mirror enclosed by the dashed circle.
    (c) Measured white light transmission spectrum of the planar cavity, as a function of the effective cavity length $L_{\textrm{e}}$. The modes are split in energy due to the birefringence of CsPbBr$_3$. The dashed green line indicates the exciton energy.
    (d) White light transmission spectra evidencing a widely tunable V-H splitting via the effective cavity length.
    (e) Momentum-resolved photoluminescence spectrum.
    Vertically and horizontally polarized lower polariton bands are labeled `V' and `H', respectively.
    For the measurements in (c,d,e), the CsPbBr$_3$ crystal is effectively embedded in a planar cavity.
    }
\end{figure*}

\noindent \large  \textbf{Results}\\ \normalsize
\noindent \textbf{Tunable birefringent perovskite cavity}\\
Figure~\ref{fig:1}(a) illustrates our system: a tunable Fabry-Pérot microcavity hosting a CsPbBr$_3$ perovskite semiconductor, all inside a closed-cycle cryostat. The cavity is made by two distributed Bragg reflector (DBR) mirrors, each mounted on four piezoelectric actuators controlling their position and orientation. The bottom mirror supports CsPbBr$_3$  crystals, synthesized via chemical vapor deposition~\cite{Zhang16} and transferred onto the mirror using thermal release tape. The top mirror has several concave features, each enabling us to make a plano-concave microcavity. The concave features were made by focused ion beam milling the glass substrate prior to deposition of the DBR~\cite{Trichet15}. Figure~\ref{fig:1}(b) shows a white light transmission image of our cavity. Further details about our setup are reported in Methods and Supplementary Figure~S1.

We first discuss the linear optics of our system when the perovskite crystal is far from all concave features and effectively embedded in a planar cavity. Figure~\ref{fig:1}(c) shows white light transmission spectra versus cavity length. For each longitudinal mode we observe two resonances, one vertically and the other horizontally polarized. Their energy splitting is determined by the CsPbBr$_3$ birefringence~\cite{Bao19, Su20}. In Supplementary Figure~S2 we present transfer-matrix calculations reproducing our experimental observations and confirming our interpretation. Calculations as a function of the perovskite crystal thickness in Supplementary Figure~S3 further demonstrate that, for the crystal in our experiments, the observed resonances correspond to exciton-polaritons due to strong exciton-photon coupling.

Figure~\ref{fig:1}(c) illustrates how the V-H splitting can be tuned via the cavity length. Cuts at three distinct cavity lengths are presented in Figure~\ref{fig:1}(d). The largest V-H splitting observed, 10.4 meV, is 300-500 times greater than in GaAs cavities~\cite{Paraiso10, Bieganska21}. Figure~\ref{fig:1}(e) illustrates the V-H splitting in momentum-resolved photoluminescence spectra measured at a fixed cavity length. The two emission bands with different dispersion relation have mutually orthogonal polarization. The bands cross at the largest in-plane momentum we can measure with our microscope objective. Similar band crossings, known as diabolic points, have recently drawn interest in connection to photonic realizations of non-Abelian gauge fields~\cite{Tercas14}, Rashba-Dresselhaus Hamiltonians~\cite{Pietka19, Li22, Liang24}, and the quantum geometric tensor~\cite{Gianfrate20, Liao21}. However, experiments have remained limited by the lack of a system combining strong birefringence and continuous wave nonlinearity. Our system achieves that combination, with the added value of tunability in the birefringence (Fig.~\ref{fig:1}(c,d)) and the nonlinearity (shown ahead). \\

\begin{figure*}[!]
	\includegraphics[width=\textwidth]{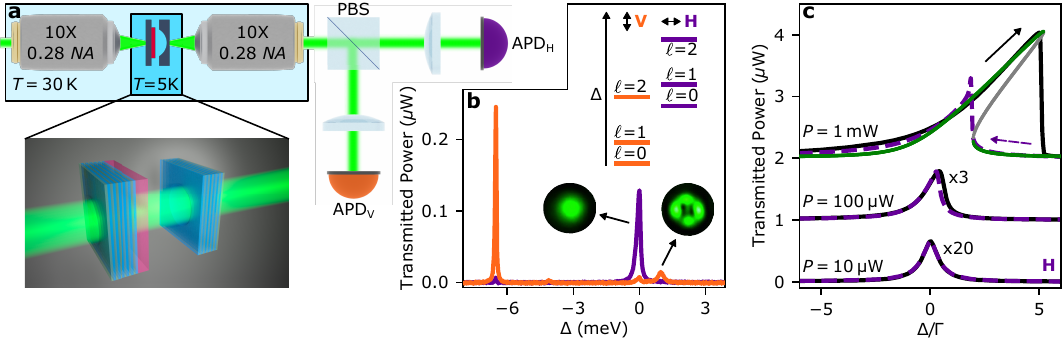}
	\caption{\label{fig:2}\textbf{Polarization modes and optical bistability.} (a) Sketch of the setup used to obtain all results in Figures~\ref{fig:2}, ~\ref{fig:3}, and ~\ref{fig:4} : A laser-driven tunable plano-concave microcavity containing a CsPbBr$_3$ crystal, all in a closed-cycle cryostat. The polarization-resolved transmitted power is measured using a polarizing beam splitter (PBS) and avalanche photodetectors APD$_H$ and APD$_V$.
    (b) Orange and purple curves are transmitted powers collected in V and H polarization, respectively. The incident laser is diagonally polarized. Its $P=100\ \mu$W power is sufficiently low to ensure linear response. $\Delta$ is the frequency detuning between the laser and the $\ell=0$ H-polarized resonance, which we control by scanning the cavity length. The spectrum is independent of scanning direction in this linear regime.
    Top inset: Schematic of the first three energy levels (transverse cavity modes) for each polarization, labeled with their angular momentum number $\ell$. Side insets: Measured transmitted intensity profiles at the indicated resonances.
    (c) Solid black and dashed purple curves are the transmitted power, averaged over 70 cycles, when scanning $\Delta$ forward and backward, respectively. $P$ is the laser power. Unlike in (b), here the incident laser is horizontally polarized. $\Gamma$ is the resonance linewidth in the linear regime.
    Green and gray curves are stable and unstable steady states of a single-mode nonlinear cavity, calculated as explained in Methods. For clarity, measurements at $P=10\  \mu$W and $P=100\  \mu$W are multiplied by 20 and 3, respectively, and measurements at $P=100\  \mu$W and $P=1$~mW are displaced vertically by $1\  \mu$W and $2\  \mu$W, respectively.
}
\end{figure*}

\noindent \textbf{CW nonlinearity}\\
We now switch to a plano-concave cavity configuration, enabling us to probe CW nonlinearities of single and coupled modes. To this end, we place a concave feature above the CsPbBr$_3$ crystal and along the optical axis. We drive the cavity with a single-mode 532 nm CW laser, and measure the polarization-resolved transmission as illustrated in Fig.~\ref{fig:2}(a). As a stepping stone, Fig.~\ref{fig:2}(b) shows transmission spectra for a relatively weak laser power 100 $\mu$W ensuring linear response. The incident light is diagonally polarized, thereby exciting both V and H polarized transverse cavity modes.

Figure~\ref{fig:2}(b) contains two insets showing how the transmitted intensity profiles resemble atomic orbitals. We label the resonances by their angular momentum number $\ell$. Notice the proximity of the $\ell=0$ H-polarized and $\ell=2$ V-polarized resonances, characterized by `s'-  and `d'-like mode profiles, respectively. Their proximity is due to the CsPbBr$_3$ birefringence, which shifts V and H resonances as illustrated in the Fig.~\ref{fig:2}(b) top inset. In an empty cavity without CsPbBr$_3$,  these two resonances would be highly detuned since H and V-polarized resonances with the same $\ell$ are degenerate. Imperfect concave mirrors with ellipticity can lift the degeneracy, but we do not observe this effect with the concave mirror under study and with our measurement resolution.

Figure~\ref{fig:2}(c) shows transmission spectra around the $\ell=0$ H-polarized resonance for three distinct laser powers $P$.  Spectra are referenced to the linear resonance linewidth $\Gamma$. The incident light is H polarized, excluding the excitation of V-polarized resonances. Solid and dashed curves correspond to cavity length scans whereby the laser-cavity detuning $\Delta$ (exact definition in Methods) increases and decreases, respectively. For $P=10\ \mu$W we observe a Lorentzian lineshape characterizing the linear regime. Around $P=100\ \mu$W the resonance bends and displays a small hysteresis. Bending towards positive  $\Delta$  indicates that polariton-polariton interactions are repulsive, or equivalently that the nonlinear refractive index $n_2$ is negative. For $P=1$ mW the hysteresis widens and we observe optical bistability: two stable states with different intensity at a single driving condition. This is the first observation of optical bistability, the hallmark of a CW nonlinearity, in a halide perovskite. \\

\begin{figure*}[!]
	\includegraphics[width=\textwidth]{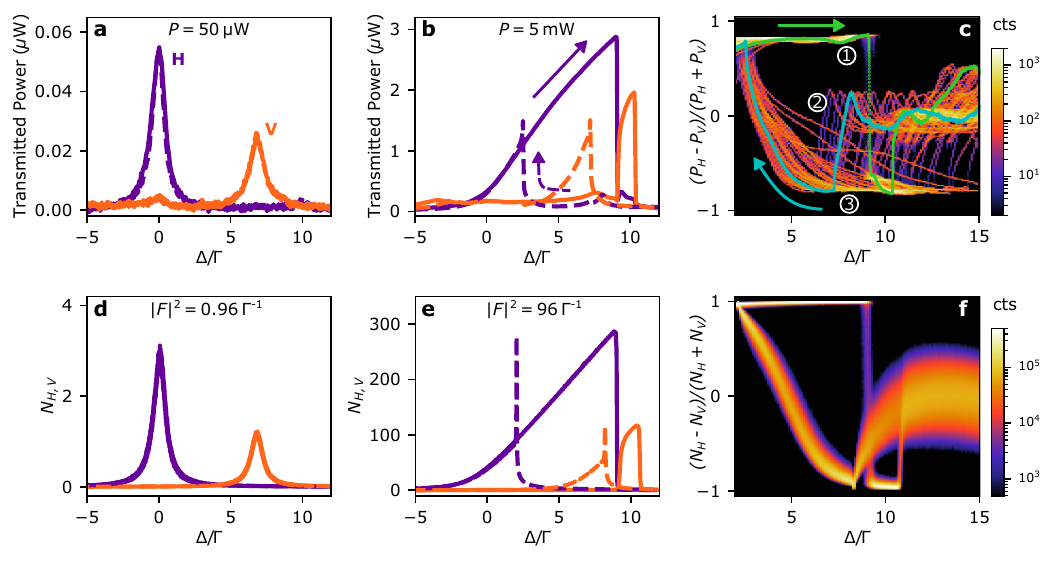}
	\caption{\label{fig:3}\textbf{Nonlinear polarization rotation and multistability.} Polarization-resolved transmitted power in the (a) linear, and (b) nonlinear, regime obtained by setting the incident power $P$ to the indicated value.  The incident laser is diagonally polarized. Purple and orange curves correspond to H and V-polarized detection, respectively.  Solid and dashed curves correspond to cavity length scans whereby $\Delta$ increases and decreases, respectively.
    (c) Histogram of the polarization contrast $(P_H-P_V)/(P_H+P_V)$ spectrum for the same driving conditions as in (b). One forward and one backward trajectory are highlighted in green and cyan, respectively. Encircled numbers label states. (d-f) Calculations corresponding to (a-c). The number of intra-cavity polaritons $N_{H,V}$ in (d,e) replaces the experimental transmitted power in (a,b). The histogram in (f) was obtained by calculating the polarization contrast spectrum for 1024 different realizations of the noise.}
\end{figure*}

\noindent \textbf{Polarization rotation and multistability}\\
We now return to diagonally polarized driving, and probe the interplay of nonlinearity and polarization in our cavity. Figures~\ref{fig:3}(a) and ~\ref{fig:3}(b) show polarization-resolved transmission spectra in the linear and nonlinear regime, respectively.  In the linear regime,  the V-polarized resonance has lower amplitude than the H-polarized resonance. While the incident polarization axis is exactly midway between the V and H crystal axes, the spatial overlap of the incident Gaussian beam with the modes is different. The overlap is smaller with the `d'-like mode profile of the V-polarized resonance, resulting in its lower excitation efficiency.  In the nonlinear regime obtained for $P=5$ mW, Fig.~\ref{fig:3}(b) shows optical hysteresis and bistability. Notice the power rise in V polarization shortly after the power drop in H polarization around $\Delta/\Gamma =9$. This behavior suggests energy transfer between H and V polarization. Such energy transfer is impossible in the linear regime, where the modes are orthogonal and decoupled. However, at sufficiently high intensities the modes couple nonlinearly as explained  ahead.

To characterize the output polarization state and its fluctuations, we plot in Fig.~\ref{fig:3}(c)  a histogram of the polarization contrast $(P_H-P_V)/(P_H+P_V)$ spectrum. $P_H$ and $P_V$ are the power in H and V polarization, respectively. Excitation conditions are the same as in Fig.~\ref{fig:3}(b). The histogram contains $40$ trajectories measured for both increasing and decreasing  $\Delta/\Gamma$; that number is limited by an uncontrolled cavity length drift in our cryostat. Starting in state `1' and increasing $\Delta/\Gamma$, the polarization state evolves along the green arrow until it rotates from  H to V around $\Delta/\Gamma =9$. The system then settles in state `3', and upon further increasing $\Delta/\Gamma$ it jumps to state `2'. State `2' is detuned from both resonances, and the powers in H and V polarization are therefore equal and low.  As a result, state `2' appears broader in Fig.~\ref{fig:3}(c) than states `1' and `3'; its mean intra-cavity intensity is low, while the noise variance is unchanged.

After reaching the largest  $\Delta/\Gamma$ in our protocol, the direction of our cavity length scan is reversed. The polarization state then returns along state `2', jumps down into state `3', and finally rises smoothly to state `1' following the curved arrow in Fig.~\ref{fig:3}(c). To illustrate this path, we highlight in cyan one trajectory measured in the backward scan. Notice how the jump from state `2' to state `3' in the backward scan (sometimes) occurs at a smaller $\Delta/\Gamma$ than the jump from state `1' to state `3' in the forward scan. This feature is indicative of tristability,  i.e., three stable states at a single driving condition. In Supplementary Figure~S4 we show, theoretically, that our system is indeed expected to be tristable in this regime.
Overall, Fig.~\ref{fig:3}(c) showcases the potential to fully rotate the polarization state of light leveraging the CW nonlinearity of our CsPbBr$_3$ cavity.

Nonlinear interaction of polarization modes has been achieved in few other experimental platforms, such as GaAs~\cite{Paraiso10} and fibre~\cite{Moroney22} cavities. These systems offer a rich playground for exploring functionalities emerging from interacting polarization modes, such as non-reciprocity~\cite{Duggan19}. In this vein, our results position CsPbBr$_3$ on par with these well-established platforms for nonlinear optics. However, as shown in the remainder of this manuscript,  the nonlinearity of our system is unique in its characteristics.

To elucidate the physics of our system and the characteristics of its nonlinearity, we developed a coupled-mode model as explained in Methods. It describes two complex-valued fields, $\alpha_{H,V}$, coherently driven with amplitudes reproducing the resonance amplitudes observed in Fig.~\ref{fig:3}(a).  The frequency detuning between the two modes matches the one observed in Fig.~\ref{fig:3}(a).
In addition, each mode is subject to dissipation, white noise, and two types of interaction. We include a self-interaction proportional to the intensity in the same mode, and a cross-interaction whereby the intensity in one mode modifies the frequency of the other and viceversa. The latter is similar to a cross-Kerr nonlinearity~\cite{Hoi13} or cross-phase modulation~\cite{Venkataraman13, Copie19}, albeit the response of our system is not instantaneous. Figures~\ref{fig:3}(d-f) show calculations based on our model, reproducing our observations in Figures~\ref{fig:3}(a-c).

The good agreement between experimental and theoretical spectra depends on several assumptions and constraints  in the model that elucidate the properties of our system.  First, self-interactions need to be non-instantaneous. Otherwise, the overshoot arising when decreasing $\Delta$ (see dashed curve in Fig.~\ref{fig:2}(c) for largest power) cannot be observed with our limited photodetection bandwidth. Indeed, from the width of that overshoot we deduce a characteristic time for the self-interactions $\tau_s=4\ \mu$s (see Supplementary Figure S5).
Second, cross-interactions also need to be non-instantaneous. Otherwise, the power in V polarization rises abruptly, rather than gradually, after the power drop in H polarization. Supplementary Figure S6 illustrates how theoretical spectra deviate from our experimental observations if self- and cross-interactions are assumed to be instantaneous. Our calculations further reveal that the cross-interaction time $\tau_c$ is similar to $\tau_s$. Third, our model enables to constraint the relative strength of self- and cross-interactions. We find that the two interactions are approximately equally strong. Deviating from this condition, even by a factor of three, results in major modifications to the calculated lineshapes (see Supplementary Figure S7). \\

\noindent \textbf{Signatures of criticality}\\
Recent experiments with CSPbBr$_3$ polaritons at room temperature displayed ultrafast nonlinearities under femtosecond-pulse excitation; these were attributed to polariton-polariton interactions~\cite{Tao22, Feng21, Peng22}. In contrast, our experiments involving CW driving display non-instantaneous interactions of a presumably different origin. The type of interactions in our model is mathematically equivalent to those  arising when an optical mode couples to an exciton reservoir ~\cite{Schmidt19, Amelio20} or a thermal field ~\cite{Geng_PRL2020, Peters21, Keijsers24}. However, unlike any other polariton or thermo-optical cavity, our CsPbBr$_3$ cavity exhibits an intriguing temperature-dependent interaction strength indicative of critical or near-critical  behavior. To evidence this behavior in a minimal configuration, we probe our cavity with H-polarized light and measure the optical hysteresis for variable temperature.

\begin{figure}[!]
	\includegraphics[width=\columnwidth]{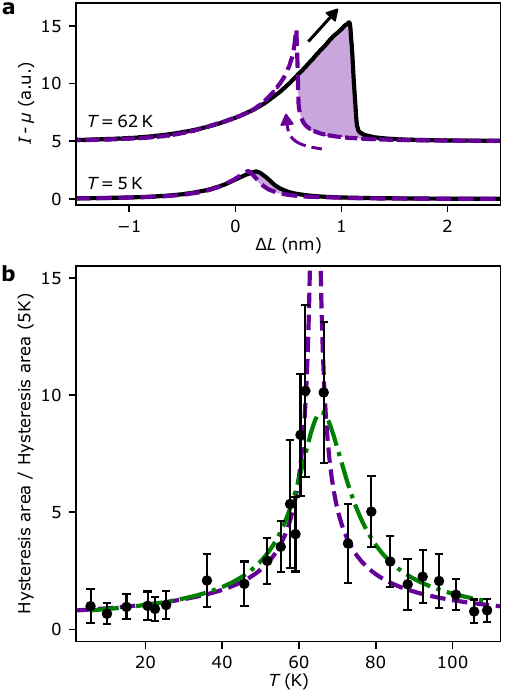}
	\caption{\label{fig:4}\textbf{Optical hysteresis as a probe for phase transitions of matter.}
    (a) Transmitted intensity, averaged over 50 cycles, for $T=5$~K and $T=62$~K. Solid black and dashed purple curves correspond to scans with increasing and decreasing cavity length $\Delta L$, respectively. The shaded region indicates the hysteresis area. The mean of the noise $\mu$ was subtracted from the transmitted intensity. For clarity, the $T=62$~K measurement is displaced vertically. (b) Average hysteresis area, referenced to the area at $T=5$ K, versus temperature. Dots and errorbars correspond to the mean and standard deviation, respectively, of 50 measurements.
    Dashed purple curve is a power-law fit to the experimental data. Dash-dotted green curve is obtained from  a fifth-order polynomial fit to the integrated hysteresis area, as explained in Supplementary Figure S8.
    The mean squared residuals of the fits corresponding to first and second order phase transitions are 0.97 and 0.61, respectively.
    For all measurements, the incident laser was H-polarized, and the power was $P=600\ \mu$W.
    }
\end{figure}

Figure~\ref{fig:4}(a) shows averaged optical hysteresis measurements at two temperatures $T$ and fixed laser power. Since we probe a single mode and everything besides temperature is unchanged, the larger hysteresis at $T=62$~K must be due to an enhanced self-interaction strength ($U_s$ in our model, described in Methods). To elucidate the origin of this enhancement, in Fig.~\ref{fig:4}(b) we plot the hysteresis area  [shaded areas in Figure~\ref{fig:4}(a)] versus temperature. The hysteresis area, and hence $U_s$,  surges as $T$ approaches $65$~K from either side.  This behavior is indicative of a phase transition in CsPbBr$_3$. To assess the plausibility of a first or second order phase transition, we fitted our data with two functions. One (dashed purple curve) fit is a power law of the form $C |T-T_c|^{-\gamma}$, with the constant $C$, critical temperature $T_c$, and exponent $\gamma$  as fit parameters. The divergence at $T_c$, consistent with out experimental data, is the hallmark of a second order phase transition. The second function we fitted  (dash-dotted green curve) corresponds to a first order phase transition, where a finite enhancement of the nonlinearity is expected; the fitting approach for this case is explained in Supplementary Figure S8. While the fit implying a second order phase transition is somewhat better (see figure caption), a first order transition cannot be excluded. Further research is needed to understand the origin of our observations, which might be related to structural transformations of CsPbBr$_3$~\cite{Rodova03, Isarov17, Li20}.

At temperatures above those considered in Fig.~\ref{fig:4}(b), we observed a significant reduction in transmitted intensity (see Supplementary Figure S9) and thus of the signal-to-noise ratio (SNR). We tentatively attribute this lower SNR to the reduction in quantum efficiency of CsPbBr$_3$ with temperature (see Supplementary Figure S10)~\cite{Li16, Shinde17}. While our experiments involve coherent instead of incoherent driving as in photoluminescence, re-absorption mechanisms degrading the quantum efficiency also degrade the transmission. Our measurements also indicate that the CW nonlinearity of CsPbBr$_3$ vanishes for  $ T \gtrsim 150$~K; see Supplementary Figure S9. This is consistent with the fact that all room-temperature CsPbBr$_3$ nonlinearities reported to date have required pulsed excitation. Clearly, the CW nonlinearity here studied is of a different origin. \\

\noindent \large  \textbf{Discussion}\\ \normalsize
To conclude, we offer three perspectives based on our results. First, the combination of CW nonlinearity and birefringence in CsPbBr$_3$ cavities is promising for exploring spin-orbit coupling physics in strongly-interacting regimes. Spin-orbit coupling phenomena in linear birefringent cavities have  received significant attention recently~\cite{Pietka19, Li22, Liang24, whittaker21, Liao21, Polimeno21, Krol21, Muszynski22, Ren22, LempickaMirek22, Su21b}. However, the effects of self- or cross-interactions on such phenomena remain largely unknown. Second, the signatures of critical or near-critical behavior presented in Fig.~\ref{fig:4} raise the interesting prospect of realizing strongly correlated states of light and matter~\cite{Cotlet15, Ashida, Bloch22} in perovskite cavities. Emergent phases of matter (e.g. ferroelectricity in CsPbBr$_3$~\cite{Li20}) could exert a major influence on light and viceversa, given the strong light-matter coupling in our system. Establishing the microscopic origin and exact strength of the CsPbBr$_3$ nonlinearity will be a key step in that direction. We note that questions about the microscopic origin and strength of nonlinear effects in other polariton systems have lingered for decades, and even resulted in claims differing by many orders of magnitude in interaction strength~\cite{Estrecho19}. For CsPbBr$_3$, we expect these questions to be settled on a shorter time scale given the much larger community of (materials) scientists working on CsPbBr$_3$. Third, we believe that CsPbBr$_3$ cavities at low temperature are promising for probing charge transport by polaritons. While polaritons are charge neutral and do not carry electrical current in conventional systems (only exciton currents arise~\cite{Feist15}), a significant polariton photocurrent has been predicted in media with broken inversion symmetry~\cite{Morimoto20}. Recent observations of the Rashba effect~\cite{Isarov17} and ferroelectricity~\cite{Li20} in CsPbBr$_3$ indicate that inversion symmetry can be indeed be broken in this material. Since  birefringence is generally associated with inversion symmetry breaking, and the birefrigence of our system also depends on temperature (see Supplementary Figure S10), CsPbBr$_3$ cavities stand as a particularly convenient platform for experiments in this third direction too.

\clearpage

\noindent \textbf{METHODS} \\
\noindent \textbf{Optical cavity}\\
Both DBRs have peak reflectance of 99.9 \% at 530 nm.
The cavity is inside a closed-cycle cryostat made by Montana Instruments, model HILA. Supplementary Information Section A includes a detailed description of our cryostat, along with a photo and a technical drawing of our setup in Supplementary Figure S1.
The measurements presented in Figs.~\ref{fig:1}(c-e), \ref{fig:2}(b,c) and \ref{fig:3}(a-c) were performed with the cryostat base temperature set to 5 K. For the measurements in Fig.~\ref{fig:4}, we attached a temperature probe to the mirror mount containing the perovksite crystal. This was done to measure the actual sample temperature, which we later found could exceed the cryostat base temperature by 1-2 K. The slightly larger sample temperature is insignificant for the previous results since, as Fig.~\ref{fig:4} shows, the interaction strength does not significantly change in the 5-7 K temperature range. Finally, we note that the cryostat pump was temporarily turned off for the measurements in Figs.~\ref{fig:1}(c,d,e). This was done  to minimize  the effects of mechanical vibrations. Those vibrations significantly impact our measurements when the cavity length is constant as in Figs.~\ref{fig:1}(e), or slowly scanned as in Figs.~\ref{fig:1}(c,d). However, in all other experiments, we circumvent the deleterious effects of vibrations by scanning the cavity length sufficiently fast. In particular, the  rate at which we sweep the laser-cavity detuning through the optical resonance is greater than the highest mechanical frequency in our setup. \\

\noindent \textbf{Cavity length calibration}\\
The horizontal axis of Fig.~\ref{fig:1}(c) was calibrated by making use of the fact that subsequent longitudinal modes are spaced by half a wavelength. Next, we considered the expression for the energy of longitudinal cavity modes: $E_{c} = hcq/2nL_e$, with $h$ Planck's constant, $c$ the speed of light in vacuum, $q$ the longitudinal mode number, $n$ the refractive index of the intra-cavity medium, and $L_e$ the effective cavity length. This enabled us to determine $q$ and $L_e$ by fitting the expression for $E_{c}$ to the experimentally observed resonances in 2.13 - 2.3 eV range. $L_e$ is an effective cavity length because it takes into account the field penetration depth into the DBRs and the optical (rather than physical) thickness of the CsPbBr$_3$ crystal. Otherwise, the above expression would only hold for a cavity made by perfectly conducting metallic mirrors and filled with a homogeneous medium.

The horizontal axis in Fig.~\ref{fig:1}(d) was obtained by subtracting the mean energy $\bar{E}$ of the two resonance peaks from the energy measured with the spectrometer.

To determine the relative cavity length $\Delta L$ in Fig.~\ref{fig:4}(a), we used a secondary laser with $632.8$ nm wavelength. Since this wavelength is outside the DBR stopband, transmission resonances are broad. This enabled us to measure the transmitted intensity of the secondary laser across a wide cavity-length range, and subsequently determine the change in cavity length $\Delta L$ as reported in Fig.~\ref{fig:4}(a). The relation between the intensity of the secondary laser and $\Delta L$ was established using the fact that consecutive longitudinal resonances are half-a-wavelength apart.\\

\noindent \textbf{Photoluminescence experiments}\\
For the photoluminescence measurements in Figs.~\ref{fig:1}(e), the cavity was illuminated by a 405 nm CW laser with $P=100\ \mu \mathrm{W}$ power. We obtained the dispersion relation by imaging the back focal plane of the collection objective onto the entrance slit of a spectrograph. \\

\noindent \textbf{Coherent driving experiments}\\
For the measurements in Figs.~\ref{fig:2}(b,c),~\ref{fig:3}(a-c) and~\ref{fig:4}, we used a concave mirror with 10 $\mu$m diameter and a 25 $\mu$m radius of curvature to make a plano-concave cavity. The polarization of the incident laser was controlled with a half-wave plate. The cavity transmission was decomposed into orthogonal linear polarizations by a polarizing beam splitter (PBS). The PBS was rotated about the optical axis to ensure its alignment with the crystalline axes of CsPbBr$_3$.
For the measurements in Figs.~\ref{fig:2}(c), \ref{fig:3}(a-c), and \ref{fig:4}, we modulated the detuning by increasing and subsequently decreasing the mirror spacing at modulation frequency $f_\mathrm{mod}=130$ Hz. \\

\noindent \textbf{Theoretical model} \\
We model our experimental system using coupled integro-differential stochastic equations for complex-valued fields $\alpha_{H}$ and $\alpha_{V}$, corresponding to the $\ell=0$ H-polarized and $\ell=2$ V-polarized modes, respectively:

\begin{subequations}
    \begin{align}
        \begin{split}
            \dot{\alpha}_H(t) =& \biggl( i\Delta - \frac{\Gamma}{2}
            - iU_s \int_0^t ds\ K_s(t-s)|\alpha_H(s)|^2 \\
            &- iU_c \int_0^t ds\ K_c(t-s)|\alpha_V(s)|^2
            \biggr) \alpha_H(t) \\
            &+ (1+\rho)[\sqrt{\kappa}F + \frac{D}{\sqrt{2}}\zeta_H(t)]
        \end{split} \\
        \begin{split}
            \dot{\alpha}_V(t) =& \biggl( i\Delta + i\delta - \frac{\Gamma}{2}
            - iU_s \int_0^t ds\ K_s(t-s)|\alpha_V(s)|^2 \\
            &- iU_c \int_0^t ds\ K_c(t-s)|\alpha_H(s)|^2
            \biggr) \alpha_V(t) \\
            &+ (1-\rho)[\sqrt{\kappa}F + \frac{D}{\sqrt{2}}\zeta_V(t)].
        \end{split}
    \end{align}
\label{eq:EOM_alpha_kernel}
\end{subequations}

\noindent $\Delta=\omega-\omega_0$ is the detuning between the laser frequency $\omega$ and the $\ell=0$ H-polarized resonance frequency $\omega_0$. $\delta$ is the frequency difference between the $\ell=0$ H-polarized and $\ell=2$ V-polarized resonances. $\Gamma$ is the total loss rate; we take it to be equal for both modes, as this matches our observations. The non-instantaneous nonlinearities are represented by the terms involving an integral. $K_s(t)$ and $K_c(t)$ are memory kernels for the self- and cross-interactions, respectively. Their properties determine to what extent the past exerts an influence over the future state of the system. $U_s$ and $U_c$ are the self- and cross-interaction strength, respectively. $F$ is the driving laser amplitude and $\sqrt{\kappa}$ is the input-output rate through the mirror on which the laser impinges. $\rho$ is the driving imbalance, determined by the incident laser polarization in our experiments. Note that, even if the incident polarization is diagonal, $\rho$ is non-zero because the overlap of the incident Gaussian beam with the two modes under consideration is different. Finally, $\zeta_{H/V}$ are stochastic processes representing Gaussian white noise in the laser amplitude and phase. They have zero mean, unit variance, and delta correlation. Moreover, all noise sources are uncorrelated. Since $\zeta_{H/V}$  are multiplied by $D/\sqrt{2}$, the standard deviation of the laser noise is $D$.

We assumed memory kernels of the form $K_s(t)=\exp{(t/\tau_s)}/\tau_s$ and $K_c(t)=\exp{(t/\tau_c)}/\tau_c$. These kernel functions are integrable and continuously differentiable. Hence, by defining $w_{s,H/V} = U_s \int_0^t ds\ K_s(t-s)|\alpha_{H/V}(s)|^2$ and $w_{c,H/V} = U_c \int_0^t ds\ K_c(t-s)|\alpha_{V/H}(s)|^2$, we can rewrite Eqs.~(\ref{eq:EOM_alpha_kernel}) as a set of ordinary differential equations in a higher dimensional phase space using Leibniz rule, to obtain
\begin{subequations}
    \begin{align}
        \dot{\alpha}_H(t) =& \left( i\Delta - \frac{\Gamma}{2} - iw_{s,H}(t) - iw_{c,H}(t) \right) \alpha_H(t) \nonumber \\
        &+ (1+\rho)[\sqrt{\kappa}F + \frac{D}{\sqrt{2}}\zeta_H(t)]  \\
        \dot{\alpha}_V(t) =& \left( i\Delta + i\delta - \frac{\Gamma}{2} - iw_{s,V}(t) - iw_{c,V}(t) \right) \alpha_V(t) \nonumber \\
        &+ (1-\rho)[\sqrt{\kappa}F + \frac{D}{\sqrt{2}}\zeta_V(t)] \\
        \dot{w}_{s,H}(t) =& \frac{1}{\tau_s} \left( U_s|\alpha_H(t)|^2 - w_{s,H}(t) \right) \\
        \dot{w}_{s,V}(t) =& \frac{1}{\tau_s} \left( U_s|\alpha_V(t)|^2 - w_{s,V}(t) \right) \\
        \dot{w}_{c,H}(t) =& \frac{1}{\tau_c} \left( U_c|\alpha_V(t)|^2 - w_{c,H}(t) \right) \\
        \dot{w}_{c,V}(t) =& \frac{1}{\tau_c} \left( U_c|\alpha_H(t)|^2 - w_{c,V}(t) \right).
    \end{align}
    \label{eq:EOM_w}
\end{subequations}
$w_{s,H/V}$ and $w_{c,H/V}$ are real-valued dynamical variables. In the case of thermo-optical cavities, they correspond to a temperature. Here, since we are unsure of the microscopic origin of the nonlinearity in CSPbBr$_3$, we can only state that they  represent the degrees of freedom which give memory to the self- and cross-interactions of the modes. $\tau_s$ and $\tau_c$ are the memory times for the self- and cross-interactions. They are the equivalent to thermal relaxation times in the case of thermo-optical cavities, and to the exciton reservoir lifetime in the case of polariton systems coupled to an exciton reservoir.

Calculations were performed by solving Eqs.~(\ref{eq:EOM_w}) using the xSPDE Matlab toolbox \cite{xSPDE}. From the fields $\alpha_{H/V}$ we computed the number of H and V-polarized polaritons $N_H=|\alpha_H|^2$ and $N_V=|\alpha_V|^2$, respectively.
We set $\kappa=\Gamma/2$, $\rho=0.22$, $\delta=-6.8\Gamma$, $U_s=\Gamma/32$, $U_c=0.9U_s$, $\tau_s=150/\Gamma$ and $\tau_c=250/\Gamma$ to reproduce our experimental observations. $F=0.98\ \sqrt{\Gamma}$ in Fig.~\ref{fig:3}(d), and $F=9.8\ \sqrt{\Gamma}$ in Figs.~\ref{fig:3}(e,f). We set $D=0.025\ \sqrt{\Gamma}$ in Figs.~\ref{fig:3}(d,e), and $D=1.25\ \sqrt{\Gamma}$ in Fig.~\ref{fig:3}(f). For all dynamical calculations, we took into account the finite bandwidth of our expeirmental APDs by downsampling numerical data.

Steady states in Fig.~\ref{fig:2}(c) were calculated by setting the time derivatives in Eqs.~(\ref{eq:EOM_w}) to zero and solving for $|\alpha_H|^2$. The stability of the steady states was determined using standard linear stability analysis, i.e. by calculating the eigenvalues of the Jacobian of Eqs.~(\ref{eq:EOM_w}).

\vspace{1cm}

\noindent \textbf{Data availability} \normalsize \\ \noindent The data in this manuscript will be uploaded to the Zenodo repository before publication. \\

\noindent \textbf{Acknowledgements} \normalsize \\ \noindent We thank Aurélien Trichet for providing the mirror sample with concave features. We thank Niels Commandeur, Ricardo Struik, and Henk-Jan Boluijt for technical support. We thank Johanna Theenhaus for initial experiments with CsPbBr$_3$ crystals.
This work is part of the research programme of the Netherlands Organisation for Scientific Research (NWO). S.R.K.R. acknowledges an ERC Starting Grant with project number 852694.  \\

\noindent  \textbf{Author contributions} \normalsize \\ \noindent G.K. and R.M.d.B. performed the experiments. Z.G. and G.K. built the setup. R.M.d.B synthesized the CsPbBr$_3$ crystals.  B.V. and K.J.H.P. developed the theoretical model. G.K. contributed to the development of the model, performed calculations, and analyzed all results together with S.R.K.R. S.R.K.R. conceived the project and supervised the work. S.R.K.R. and G.K. wrote the manuscript. All authors discussed the results and the manuscript.  \\

\noindent \textbf{Competing interests} \normalsize \\ \noindent The authors declare no competing interests.\\

\renewcommand{\thefigure}{S\arabic{figure}}
\setcounter{figure}{0}
\hyphenation{micro-cavity}

\clearpage

\begin{center}
\begin{Large}
\textbf{Supplemental Material}
\end{Large}
\end{center}
\section{Tunable cavity in a closed-cycle cryostat}

\begin{figure*}[!]
	\includegraphics[width=\textwidth]{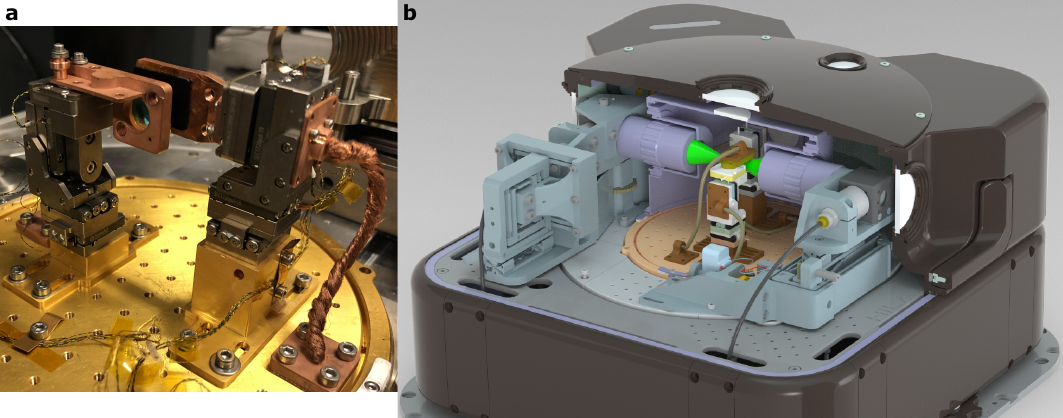}
	\caption{\label{fig:S1}\textbf{Tunable cavity in a closed-cycle cryostat.}
    (a) Photograph of our tunable cavity setup, and (b) illustration of our closed-cycle cryostat, both as described in the text.
    }
\end{figure*}

Figure~\ref{fig:S1}(a) shows a photograph of our tunable cavity setup in a closed-cycle cryostat. The DBR mirrors making the cavity are glued onto the two copper plates, each mounted on a stack of four piezoelectric actuators used to control the position and orientation of the mirrors. The actuators stand on custom-made pedestals, elevating the optical axis to the center of the cryostat's windows.

Figure~\ref{fig:S1}(b) illustrates our closed-cycle cryostat, including various custom-made parts for optical experiments. The cryostat is made by Montana Instruments, model HILA.  The cavity setup is on the orange disk at the center of the cryostat.  There, the base temperature can reach $4.2$~K at maximum cooling power. The sample is cooled through thermal links connected to the base. A custom-made radiation shield with intrusions encapsulates the cavity setup while allowing two microscope objectives to approach the sample. This allows us to measure in transmission with the objectives outside of the radiation shield. Custom-built mounts for the microscope objectives are placed on the gray annulus around the central disk. The platform comprising the central disk and annulus is actively stabilized by electromagnetic springs to dampen mechanical vibrations. Moreover, the platform is designed to have a large mass, endowing the platform with higher inertia and greater stability than standard cryostats.

\section{Transfer matrix calculations}

\begin{figure*}[!]
	\includegraphics[width=\textwidth]{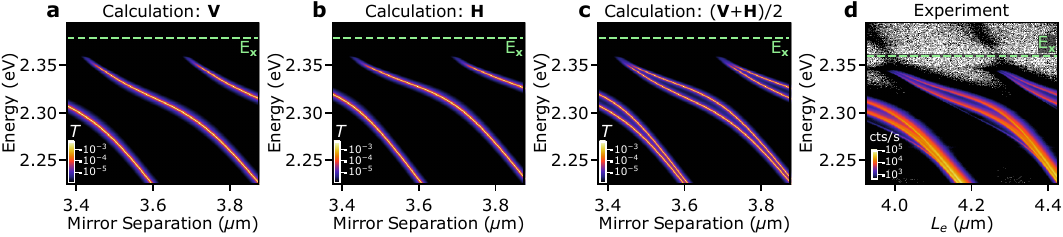}
	\caption{\label{fig:S2}\textbf{Transfer matrix calculations reproducing experimental observations.}  All figures show transmittance spectra as a function of cavity length.
    (a-c) Transfer matrix calculations for CsPbBr$_3$ layer thickness $L_c=275$ nm. (a) Calculated transmittance for V-polarized illumination, using only $n_V$. (b) Calculated transmittance for H-polarized illumination, using only $n_H$. (c) Transmittance for unpolarized illumination, obtained from the average of the spectra in (a) and (b).
    (d) Experimental white light transmission spectrum of the planar cavity, as function of the effective cavity length $L_e$. (d) shows the same data as Fig.~1(c) of the main text.
    Green dashed lines indicate exciton energies.
    }
\end{figure*}

Here we present transfer matrix calculations reproducing the measurements in Figure~1(c) of the main text, and supporting our interpretation of the resonances observed therein. We consider a DBR-CsPbBr$_3$-vacuum-DBR cavity. The bottom and top DBR consist of 10 and 11 pairs of layers, respectively. The refractive indices of the layers are $n=1.46$ and $n=2.15$. These refractive index values correspond to silicon and Ta$_2$O$_5$, as used in the experiment, in the relevant spectral range. We set the CsPbBr$_3$ layer thickness to $L_c=275$ nm, corresponding to our experiments. The refractive index of CsPbBr$_3$ was determined from absorption measurements not shown. Those measurements involved a nominally identical but in-practice different CsPbBr$_3$ crystal to the one in our cavity experiments. We   used a different crystal  because we needed to place it on  a transparent substrate for the absorption measurements. We determined the imaginary part of the frequency-dependent refractive index from the absorption spectrum, and the corresponding real part of the refractive index via the Kramers-Kronig relations as done in Ref.~\onlinecite{Geng21} for similar  CsPbBr$_3$ crystals. To set the refractive index at energies away from the exciton resonance, we added background refractive indices $n_H=2.56$ and $n_V=2.64$, following Ref.~\citealp{Bao19}. Here, $n_H$ and $n_V$ were used to calculate the transmittance spectra for horizontally and vertically polarized illumination, respectively. We calculated the transmittance at normal incidence for the two polarizations separately. To account for the unpolarized illumination in our experiment, the two sets of linearly polarized transmittance spectra were incoherently averaged. Finally, to vary the cavity length we varied the thickness of the vacuum layer. Note that the total separation between the DBRs includes both the thickness of the vacuum layer and the physical thickness of the CsPbBr$_3$ layer.

Figure~\ref{fig:S2}(a) and~\ref{fig:S2}(b) show the calculated transmittance for V-polarized and H-polarized illumination, respectively. Both calculations show a single resonance per longitudinal mode. The averaged transmittance for both polarizations is shown in Fig.~\ref{fig:S2}(c). It shows two resonances per  longitudinal mode, as in our experiments. The splitting between the two resonances is caused by the difference between $n_V$ and $n_H$, due to the CsPbBr$_3$ birefringence. Fig.~\ref{fig:S2}(d) shows the same white light transmission measurements in Fig.~1(c) of the main text, here reproduced to facilitate comparison to the calculations. Clearly, the calculations in Fig.~\ref{fig:S2}(c) reproduce the main observations in Fig.~\ref{fig:S2}(d).

Differences in the horizontal axes of Fig.~\ref{fig:S2}(c)  and ~\ref{fig:S2}(d) are due to two effects combined: The field penetrates into the DBRs, and the optical thickness (refractive index times thickness) of the perovskite crystal is over twice its physical thickness. Hence the effective cavity length is greater than the mirror separation. Note also that the exciton energy in the calculations of Fig.~\ref{fig:S2}(c) deviates slightly from the experimental one in Fig.~\ref{fig:S2}(d). This deviation is due to the fact that we used a different crystal for the cavity experiments and for the absorption measurements from which the refractive index was obtained.  As a result of this sample-to-sample difference, the exciton-photon anti-crossing is shifted to higher energy in the calculations. Finally, we note that resonance linewidths are smaller in theory than in experiments. This is due to the fact that the calculation involves normal incidence illumination only, whereas the experiment involves a distribution of illumination angles~\cite{Geng21}. The width of that distribution is determined by the $0.28$ numerical aperture of the microscope objectives we used.

\begin{figure*}[!]
	\includegraphics[width=\textwidth]{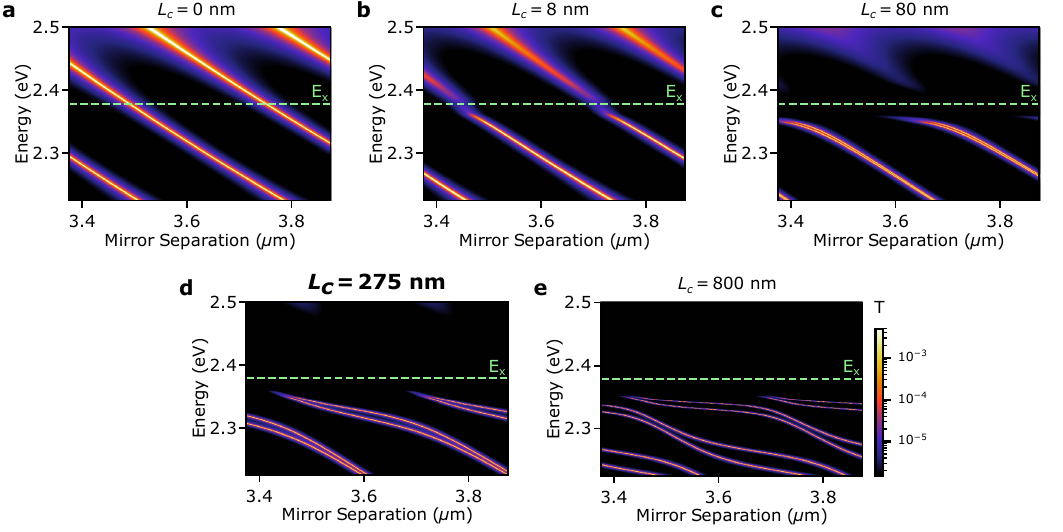}
	\caption{\label{fig:S3}\textbf{Transmission spectra for various crystal thicknesses.}
    Transmittance as a function of cavity mirror separation, obtained from transfer matrix calculations. The thickness $L_c$ of the CsPbBr$_3$ layer is indicated at the top of each panel. Green dashed lines indicate the exciton energy. (a) corresponds to a bare cavity without CsPbBr$_3$ crystal. The calculations in (d) correspond to Fig.~\ref{fig:S2}(c) and reproduce the experimental transmission spectra. All panels are plotted on the same colorscale.}
\end{figure*}

Next we support our claim in the main text that the observed resonances correspond to polaritons.  To this end, we performed transfer matrix calculations for several values of the crystal thickness $L_c$. The results are presented in Fig.~\ref{fig:S3}. Figure~\ref{fig:S3}(a) shows the transmittance spectrum for $L_c=0$, meaning no crystal. We observe several resonances, each corresponding to a different longitudinal mode. There is no bending at the exciton energy, no polarization splitting, and no absorption above the exciton energy. At higher energies, the resonances broaden as they approach the edge of the DBR stopband where the reflectivity drops.

In the presence of a thin crystal, as shown for $L_c=8$ nm in Fig.~\ref{fig:S3}(b), the resonances broaden. The system remains in the weak coupling regime. New (polariton) modes are not formed. Next, for a thicker crystal, as shown for $L_c=80$ nm in Fig.~\ref{fig:S3}(c), we observe an avoided resonance crossing around the exciton energy, as expected for strong coupling. The $\Omega = 26$~meV Rabi splitting between upper and lower polaritons exceeds the cavity and exciton linewidths, thereby confirming strong coupling. Notice the larger linewidth and lower transmittance of the upper polariton, relative to the lower polariton. This effect is due to high-frequency absorption in the crystal, which becomes increasingly pronounced as the crystal thickness increases.

Next, Fig.~\ref{fig:S3}(d) shows the transmittance for $L_c=275$ nm, which corresponds to Fig.~\ref{fig:S2}(c) and to the experiment. Unlike in Fig.~\ref{fig:S3}(c), in Fig.~\ref{fig:S3}(d) only the lower polariton can be observed. The upper polariton has been blurred by high-frequency absorption in this thicker crystal. Nonetheless, despite the absence of a clear upper polariton feature, the bending and anti-crossing of the lower polariton as it approaches the exciton energy demonstrates that the system is in the strong coupling regime.

Finally, Fig.~\ref{fig:S3}(e) shows transmittance spectra for a much thicker crystal with $L_c=800$ nm. Similarly to Fig.~\ref{fig:S3}(d), the upper polariton is blurred by high-frequency absorption.  We furthermore observe a `waviness' of the resonances as a function of cavity length. This behavior arises because, as the cavity length changes, the electric field is more or less  concentrated in the thick crystal or in the air gap~\cite{Janitz15, Bogdanovic17}.  When the electric field amplitude is more concentrated in the crystal, the V-H splitting is larger. Thus, a judiciously selected combination of crystal thickness and cavity length can be leveraged to further enhance the polarization splitting of the resonances.


\section{Tristability}

\begin{figure}[!]
	\includegraphics[width=\columnwidth]{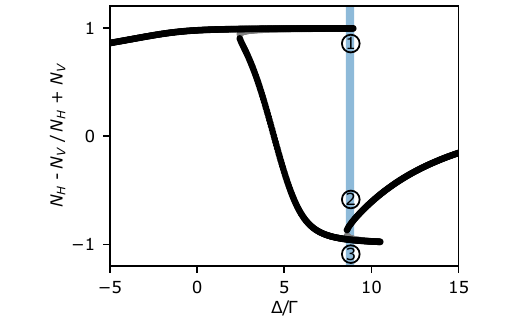}
	\caption{\label{fig:S4}\textbf{Steady state calculation evidencing tristability.}
    Polarization contrast $(N_H - N_V)/(N_H + N_V)$ of the calculated steady states as a function of detuning. The system is tristable inside the blue shaded area. Encircled numbers label states, as in Fig.~3(c) of the main text. Black and grey points indicate stable and unstable steady states, respectively.}
\end{figure}

Figure~3(c) of the main text shows signatures of tristability: three stable steady states at a single driving condition. Here we present steady-state calculations for our system, and demonstrate that tristability is indeed expected in the relevant frequency range. Figure~\ref{fig:S4} shows the polarization contrast $(N_H - N_V)/(N_H + N_V)$, as studied in the main text, versus the laser-cavity detuning $\Delta$ referenced to the total loss rate $\Gamma$. It shows tristability in the narrow $\Delta/\Gamma$ range shaded in blue. The three stable states are labeled `1', `2' and `3', as in Fig.~3(c) of the main text.

\section{Characteristic time of self-interactions}

\begin{figure}[!]
	\includegraphics[width=\columnwidth]{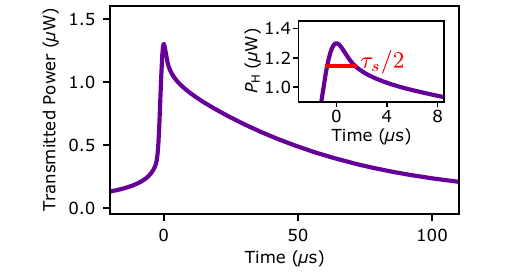}
	\caption{\label{fig:S5}\textbf{Characteristic time of self-interactions.}
    Experimental transmitted power, averaged over 70 scans, when scanning $\Delta$ backward, as a function of time. Incident laser is horizontally polarized. Data corresponds to $P=\ 1$ mW dashed purple curve in Fig.~2(c) of the main text.
    Inset: zoom of the overshoot, with FWHM indicated in red.
    }
\end{figure}

Here we determine the characteristic time of self-interactions by analyzing the measurements for $P=\ 1$ mW in Fig.~2(c) of the main text. Figure~\ref{fig:S5} shows those measurements, plotted as a function of time instead of detuning. The inset shows the overshoot, with its full width at half maximum (FWHM) highlighted in red.  For a single-mode cavity with memory as considered here, Ref.~\onlinecite{Peters22} established that the FWHM of the overshoot corresponds to half of the memory time. Thus, we conclude that self-interactions in our perovskite cavity have a characteristic (memory) time $\tau_s=4\ \mu$s.

\section{Non-instantaneous self- and cross-interactions}

\begin{figure}[!]
	\includegraphics[width=\columnwidth]{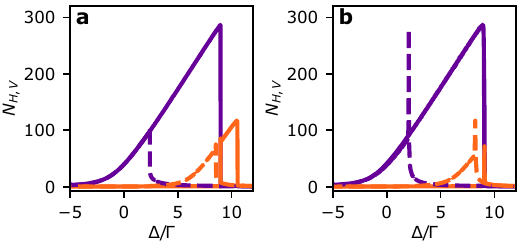}
	\caption{\label{fig:S6}\textbf{Effects of self- and/or cross-interactions being instantaneous.}
    Polarization-resolved calculated number of polaritons while varying the detuning, for different types of interactions.
    In (a), both self- and cross-interactions are instantaneous.
    In (b), self-interactions are non-instantaneous, while cross-interactions are instantaneous.
    An increasing and decreasing scan in detuning are shown in solid and dashed curves, respectively. Horizontally and vertically polarized polariton number are shown in purple and orange curves, respectively.
    }
\end{figure}

Here we demonstrate that both self- and cross-interactions need to be non-instantaneous (as implemented in our model) in order to accurately reproduce the measurements reported in Fig. 3 of the main text. To this end, we calculate spectra using models with different types of interactions, but all other simulation parameters the same as in Fig.~3(e) of the main text.
Figure~\ref{fig:S6}(a) presents calculations assuming both self- and cross-interactions are  instantaneous. This is equivalent to letting $\tau_s \rightarrow 0$ and $\tau_c\rightarrow 0$ in the model reported in Methods. Then, in Fig.~\ref{fig:S6}(b), only the cross-interactions are instantaneous; self-interactions have a finite response time.

Clearly, none of the calculations in Fig.~\ref{fig:S6} reproduce the experimental observations as good as the calculations in Fig. 3(e) of the main text. In Fig.~\ref{fig:S6}(a) the overshoots expected when decreasing $\Delta$ (dashed curves) do not  arise. Then, in  Fig.~\ref{fig:S6}(b) those overshoots arise. However, when increasing $\Delta$ (solid curves) the gradual power buildup in V-polarization is not reproduced. In fact, $N_V$ only briefly increases at the moment $N_H$ drops, and $N_V$ is no longer excited as $\Delta$ increases further. Only if both interactions are non-instantaneous, as in Fig.~3(e) of the main text, are the measurements accurately reproduced. Note that our calculations take into account the finite detection bandwidth in experiments, as mentioned in Methods.

\section{Relative strength of the self- and cross-interactions}

\begin{figure}[!]
	\includegraphics[width=\columnwidth]{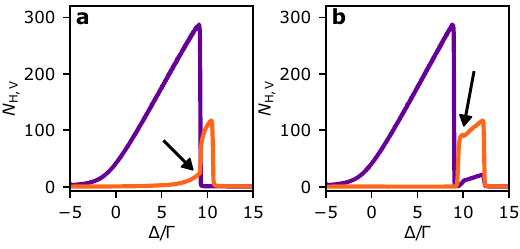}
	\caption{\label{fig:S7}\textbf{Relative strength of the self- and cross-interactions.}
    Polarization-resolved calculated number of polaritons while increasing the detuning, for different relative strengths of the self- and cross-interactions. $U_c/U_s=0.3$ in (a) and $U_c/U_s=2.7$ in (b). Arrows indicate features that are absent in the experimental spectra. Horizontally and vertically polarized spectra are shown in purple and orange, respectively.
    }
\end{figure}

Here we demonstrate how the spectrum changes when self- and cross-interaction strengths, $U_s$ and $U_c$, deviate from the values reported in Methods. Concretely, we adjust the ratio $U_c/U_s$ and keep all other simulation parameters the same as in Fig.~3(e) of the main text. We let $U_c/U_s=0.3$ in Fig.~\ref{fig:S7}(a), and $U_c/U_s=2.7$ in Fig.~\ref{fig:S7}(b); these values are 3 times smaller and greater than in Fig.~3(e), respectively. In both cases, calculated spectra exhibit features (indicated by arrows in Fig.~\ref{fig:S7}), that are absent in the experimental results. We therefore conclude that self- and cross-interactions are roughly of the same magnitude.  By `roughly' we mean that the ratio $U_c/U_s$ can be adjusted by a few percent without noticeable difference in the spectra. However, a factor of 3 or more change in $U_c/U_s$ results in qualitative discrepancies with experimental results.

\section{Polynomial fit hysteresis area}

\begin{figure*}[!]
	\includegraphics[width=\textwidth]{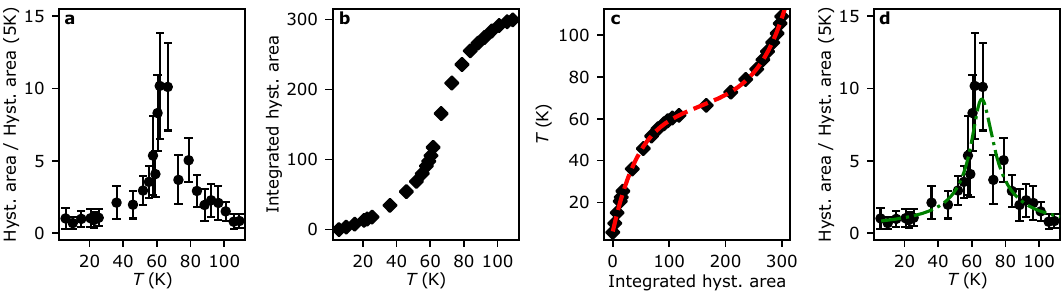}
	\caption{\label{fig:S8}\textbf{Fitting the temperature-dependent hysteresis area.}
    (a) Hysteresis area as function of sample temperature, referenced to its value at $5$ K. (b) Integrated hysteresis area as function of temperature. (c) Temperature as function of the integrated hysteresis area. Dashed red curve is a fifth-order polynomial fit to the data. (d) Same as (a), together with the inverse of the derivative of the fifth order polynomial fit of (c).
    }
\end{figure*}

Here we describe the procedure to obtain the polynomial fit of the hysteresis area, shown as a dash-dotted green curve in Fig. 4(b) of the main text.
We started from the hysteresis area, referenced to its value at $5$ K, as function of sample temperature $T$ [see Fig.~\ref{fig:S8}(a)]. The hysteresis area was integrated over temperature to obtain $A_\mathrm{int}$, which is shown in Fig.~\ref{fig:S8}(b). Next, we converted the integrated hysteresis area as function of temperature $A_\mathrm{int}(T)$ in Fig.~\ref{fig:S8}(b) to the temperature as function of the integrated hysteresis area $T(A_\mathrm{int})$ in Fig.~\ref{fig:S8}(c) by exchanging the horizontal and vertical axes. We then fit a fifth-order polynomial to $T(A_\mathrm{int})$, shown as dashed red curve in Fig.~\ref{fig:S8}(c). From the polynomial fit, we deduced the derivative $dT/dA_\mathrm{int}$.
Finally, we obtained the green curve in Fig. 4(b) of the main text (reproduced in Fig.~\ref{fig:S8}(d)) by plotting $(dT/dA_\mathrm{int})^{-1}$ as function of $T$.

\section{Vanishing transmission at high temperature}

\begin{figure}[!]
	\includegraphics[width=\columnwidth]{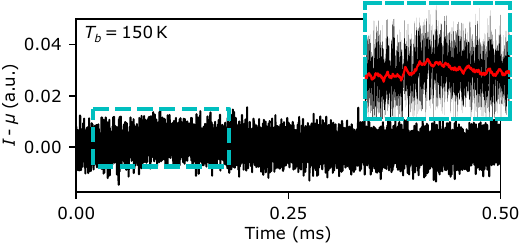}
	\caption{\label{fig:S9} \textbf{Reduced SNR at higher temperature.} Transmitted intensity as a function of time during a single forward scan in $\Delta L$, at cryostat base temperature $T_b=150$~K. The inset zooms into the region enclosed by the dashed rectangle, and includes a moving average in red. The incident laser power was $P=2$~mW, and the mean of the noise $\mu$ was subtracted from the transmitted intensity.
    }
\end{figure}

Here we report how the transmission under resonant excitation changes at a higher temperature. Figure~\ref{fig:S9} shows the transmitted intensity at the base temperature of the cryostat $T_b=150$~K during a single forward scan in the cavity length $\Delta L$. As explained in Methods, $T_b$ can be 1-2K lower than the sample temperature. cNotice the much lower signal-to-noise ratio (SNR) than in Figs.~2(c) and 4(a) of the main text, despite the larger laser power $P=2$~mW used in Fig.~\ref{fig:S9}. The transmitted intensity is plotted as function of time because the low SNR makes it difficult to determine the cavity resonance frequency. A close look at the moving mean of the transmission (see Fig.~\ref{fig:S9} inset) does reveal remnants of a resonance. Its lineshape is approximately symmetric as expected for a linear cavity. The nonlinearity therefore seems absent at $T_b = 150$~K.\\

\section{Temperature-dependent Photoluminescence}

\begin{figure}[!]
	\includegraphics[width=\columnwidth]{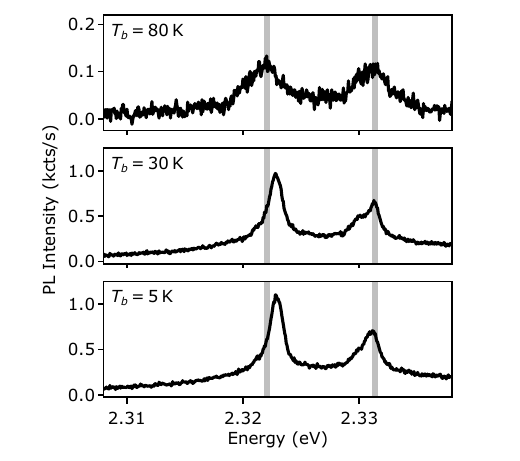}
	\caption{\label{fig:S10}\textbf{Temperature-dependent photoluminescence spectra.} Photoluminescence spectrum at $k_\parallel=0$ at three cryostat base temperatures $T_b$. Vertical gray lines indicate peak energies at $T_b=80$ K.
    }
\end{figure}

Here we report how the photoluminescence spectra changes with temperature. Figure~\ref{fig:S10} shows photoluminescence spectra at cryostat base temperatures $T_b=80$ K, at $T_b=30$ K and at $T_b=5$ K, the last one corresponding to  Figure~1(e) of the main text. The spectra were obtained from momentum-resolved measurements by integrating the dispersion over $\Delta k=0.5\ \mu \mathrm{m}^{-1}$ around $k_\parallel=0$.  Notice how peak polariton intensities at 5 K are much greater than at 80 K. This is due to the rise in quantum efficiency of CsPbBr$_3$ with decreasing temperature~\cite{Li16, Shinde17}.
We estimate a drop of roughly three orders of magnitude  in the internal quantum efficiency of CsPbBr$_3$ when increasing the temperature from $T_b = 5$~K to $T_b = 100$~K. Our estimate is based on the photoluminescence intensity measurements presented in Fig.~\ref{fig:S10}, combined with the enhanced absorption that follows from the reduced transmission in Fig.~\ref{fig:S9}.

Figure~\ref{fig:S10} also shows that it is possible to finely tune the magnitude of the V-H splitting by adjusting the temperature.  Vertical gray bars are included in the figure to facilitate comparison of the peak energies for different temperatures. The measurements show how the splitting, given by the energy difference of the two peaks in the spectra in Fig.~\ref{fig:S10}, decreases weakly with decreasing temperature.

\end{document}